\documentclass[11pt]{article}
\usepackage{graphicx}
\usepackage{epstopdf}
\usepackage{amsmath}
\usepackage{amsfonts}
\usepackage{amssymb}
\usepackage{latexsym}
\usepackage{hyperref}
\usepackage[english]{babel}
\usepackage[utf8x]{inputenc}
\usepackage{xcolor}
\usepackage{slashed}
\usepackage{bm}  
\usepackage{subfigure}
\usepackage[section]{placeins}

\begin{document}
\title{Variational Procedure for Higher-Derivative Mechanical Models in a Fractional Integral Framework}         
\author{C. F. L. Godinho$^{a}$  \footnote{crgodinho@gmail.com},\,\,\,\,Nelson Panza$^{b}$ \footnote{nelsonpanza@gmail.com} and J. A. Helay\"{e}l Neto$^{c}$  \footnote{josehelayel@gmail.com}\\\\
$^{a}$ Departamento de F\'isica,Universidade Federal Rural do Rio de Janeiro (UFRRJ),\\BR 465-07, 23890-971, Serop\'edica, RJ, Brazil.\\\\
$^{b}$ Department of Physics, Centro Federal de Educa\c{c}\~ao Tecnol\'ogica Celso Suckow da Fonseca,\\ Av Maracan\~a 229, 20271-110, Rio de Janeiro, RJ, Brazil. \\\\
$^{c}$ Centro Brasileiro de Pesquisas F\'isicas (CBPF),\\Rua Dr. Xavier Sigaud 150, Urca,22290-180, Rio de Janeiro, Brazil.}  


\date{\today}          
\maketitle
\begin{abstract}
\noindent We present both the Lagrangian and Hamiltonian procedures for treating higher-order equations of motion for mechanical models by adopting the Riemann-Liouville Fractional integral to describe their action. We point out and discuss its efficacy and difficulties. We also present the physical and geometric interpretations for the approach we pursue and present the details of a higher-order harmonic oscillator.   
\end{abstract}
\newpage
\section{Introduction}       

The study of non-linear dynamics constitutes today an important topic in the description of diverse physical and mathematical systems Its actual success and a radically new understanding of non-linear processes have however taken place over the past four decades. This understanding has been inspired by the discovery and insight of a new phenomenon known as dynamical chaos. It is easy to understand the reasons for the flourishing of the area, since any typical system with more than one degree of freedom may exhibit chaotic motion for some initial conditions. We do not know yet what the measure of chaotic trajectories is, but it seems it is non-trivial, and this makes the study of chaos of primary importance for building up models of dynamical processes in Nature \cite{GZ}. At a very fundamental level, processes of Nature are strongly dominated by non-linear effects. 

Fractional Calculus (FC) is one of the possible generalizations of classical calculus.  It has been used in several fields of science. FC provides an interesting, and sometimes unexpected, redefinition of the mathematical tools, and it seems very useful to approach anomalous and frictional systems; in particular, we can cite the continuous time random walk scheme as an illustrative physical example, where, within the fractional approach, it is possible to include external "fields"in a straightforward manner. Also, the consideration of transport in the phase space spanned by both position and velocity coordinates is possible within the same formalism. Moreover, the calculation of boundary value problems is analogous to the procedure adopted to study the corresponding standard equations \cite{klafter,9,scalas,Hilfer2}. Other important applications may be found by investigating response functions, where a great deal of problems has been reported on the phenomenon of non-exponential and power-law relaxation, which is typically observed in complex systems such as dielectric and ferroelectric materials, polymers and others. The main feature of such systems is a strong (in general, randomic) interaction between their components in the passage to a state of equilibrium. The FC approach to describe dynamical processes in disordered or complex systems, such as relaxation or dielectric behavior in polymers or photo bleaching recovery in biologic membranes, has proved to be an extraordinarily successful tool. Some authors have proposed different fractional relaxation models to study filled polymer networks and investigate the dependence of the decisive occurring parameters on the filler content \cite{stanisvasky,Metzler}.  The investigation of exactly solvable fractional models of linear viscoelastic behavior is another successful field of application.  In recent years, both phenomenological and molecular-based theories for the study of polymers and other viscoelastic materials came up with integral or differential equations of fractional order. Some current models of viscoelasticity based on FC are usually derived from the Maxwell model by replacing the first-order derivative $({d/{dt}})$
by its corresponding fractional version $({d^{\alpha}/{dt^{\alpha}}})$ \cite{Glockle}, where $\alpha$ is non-integer.

  The study of systems where the Lagrangians contain higher derivatives were long ago proposed \cite{Podol}. Models with higher derivatives, for example, Lagrangian models where $L(q^{(3)},\ddot{q},\dot{q},q)$ or cases of even higher derivatives may appear in many different areas of Physics.  In a first moment, their interest may appear to be merely academic, since most of the Lagrangian models in Physics are based only on first derivatives. It is well known that most of important physical theories are bounded to second order differential equations.  On the other hand theories and models described by high-order derivative Lagrangians appears to have great importance too, for instance, \cite{pais,Podol}.  Such theories exhibit interesting properties, including the energy unboundness from below at classical level and improving the convergence of the Feynman graphs \cite{thi}.  However the higher derivatives models impose some undesirable consequences on its dynamics stability at classical leve, since that the Noether energy is unbounded.  At the quantum level, the instability engenders ghost poles in the propagator and subsequently problems with energy spectrum.
	
The outline of our contribution is as follows: in Section 2, we present a brief recall of the variational principle using fractional calculus.  We work with the usual Riemann-Liouville integral and we revisit the well-known (FALVA) approach \cite{Rami}. Next, in Section 3, we extend the problem to include higher-order derivative systems, for which Euler-Lagrange and Hamilton equations are worked out. In Section, 4 we study a Pais-Uhlenbeck-type oscillator in the FALVA framework and identify how the fractional approach affects the results. Finally, in Section 5, we cast our Discussion and Final Considerations.

\section{The Modified Variational Principle}

Nowadays, the inter- and multi-disciplinarity among areas is an important issue, and this attitude can be quite useful to study several problems of different areas (besides the ones mentioned in the previous Section) of science such as  viscoelasticity and damping, glassy condensation, diffusion and wave propagation, electromagnetism, chaos and fractals, heat transfer, biology, electronics, signal processing, robotics, system identification, genetic algorithms, percolation, modeling and identification, telecommunications, chemistry, irreversibility, control systems as well as engineering and economy including finance \cite{ten,ten2}.

It is also well-known, from the current literature, that the fractional approach can describe more precisely a plethora of physical systems. The formalism can be applied in many classical and quantum systems as previously cited.  We areconvinced it may well be considered to investigate field-theoretic models in general.

The generalization of the concept of derivative  with non-integer values goes back to the beginning of the theory of differential calculus. Nevertheless, the development of the theory of FC is due to contributions of many mathematicians such as Euler, Liouville, Riemann, and Letnikov \cite{Old,Mill,Pod,kilbas}.

Since 1931, when Bauer \cite{Bau} showed that we cannot use the variational principle to obtain a single linear dissipative equation of motion with constant coefficients, a new horizon of possibilities was glimpsed.  Nowadays, it has been observed that bothe in Physics and Mathematics the methodology necessary to understand new questions has changed towards more compact notations and powerful nonlinear and qualitative methods. Derivatives and integrals of fractional order have been adopted in many physical applications.  For instance, questions about viscoelasticity and diffusion process may have a more detailed description when this approach is used. In Nature, the majority of systems contains an internal damping process and the traditional approach based on energy aspects cannot be used everywhere to obtain the right equations of motion and dynamical properties.

So, after Bauer's corollary, Bateman \cite{Bat} proposed a procedure where multiple equations were obtained following from a Lagrangian.  Riewe \cite{Rie} observed that using FC it was possible to obtain a formalism which could be used to describe both conservative and nonconservative systems.  Namely, using this approach one can obtain the Lagrangian and Hamiltonian equations of motion also for nonconservative systems. In \cite{Ag} Agrawal studied a fractional variational problem.  A fractal concept applied to quantum physics has been investigated and reported in \cite{Las}.

This subject has recently been re-assessed in \cite{Dre} and the solution of a fractional Dirac equation (order $\alpha\,=\,2/3$) was introduced in \cite{Ras}.

\bigskip
\subsection{Modified Equations:}

Let us consider $\eta$ as a positive number and let $f$ be a continuous function on $[0,\eta]$, then if $\alpha \ge 1$,

\begin{equation}
\int_0^t (t-\tau)^{\alpha-1} f(\tau) d \tau
\end{equation}
exists as a Riemann integral for all $t \in [0,\eta]$.
Let $Re\, \nu >0$ and let $f$ be piecewise continuous on $I=(0,\infty)$ and integrable on any finite subinterval of $I=[0,\infty)$, then for
any $t>0$ we can write
\begin{equation}
I^{\alpha}(t)={1\over \Gamma(\alpha)}\int_0^t (t-\tau)^{\alpha-1} f(\tau) d \tau
\end{equation}
as the left-sided Riemann-Liouville fractional integral of order $\alpha$.  

Fractional integral and fractional differentiation are generalizations of usual integer-order integral and  differential calculus.  However, there is some cloud on its physical and geometrical interpretations.  Some papers try to connect the fractional integral and derivatives with the fractal world, however it was shown that is an equi\-vocal \cite{Ru1,Ru2}.

It is well-known that seve\-ral definitions of fractional derivative and integral exist, for instance, Grunwald-Letnikov, Caputo, Weyl, Feller, Erdelyi-Kober and Riesz fractional derivatives as well as fractional Liouville operators which have been popularized when fractional integration is performed in dynamical systems under study \cite{Rami}, following this idea, let us consider a smooth Lagrangian function.  For any smooth path $q:[a,b]\rightarrow M$ satisfying boundary conditions $q(1)=q_1$ and $q(2)=q_2,$ we define the fractional action integral by 
\begin{equation}
\label{1}
S^{\alpha}[q](t)\,=\,{1\over \Gamma(\alpha_i)} \int_{\tau} L(q(\tau),\dot q(\tau),\tau)(t-\tau)^{\alpha-1}d\tau ,
\end{equation}
where $\Gamma(\alpha_i)$ is the traditional Euler gamma function, with $\alpha \in (0,1]$ and $\dot q\,=\,{dq/{d\tau}}$ is the derivative with respect to the intrinsic time $\tau \in (a,t^{\prime})$  and $t \in [t_0,t^{\prime}]$ is the time for some observer in a particular referential. 

Notice that the Lagrangian in Eq. (\ref{1}) is weighted by 
\begin{equation}
{1\over \Gamma(\alpha_i)}(t-\tau)^{\alpha-1} \nonumber.
\end{equation}
The time weighting acts as a time-dependent factor \cite{Ras},  
and obviously when $\alpha\rightarrow 1$ we re obtain the usual functional 
\begin{equation}
S[q]\,=\, \int_{\tau} L(q(\tau),\dot q(\tau),\tau) d\tau .
\end{equation} 
\paragraph*{} The explicit time-dependence introduced by the weight in the action that governs the dynamics in the regime of fractionality ($\alpha<1$) has the relevant consequence of violating the conservation of the energy of the system, which becomes clear if we invoke the Noether's theorem. This means that, in the fractional scenario, the systems behaves as a non-isolated system; the fractionality is parametrising some sort of interaction of the system with a external source, whose microscopic details we do not know. There is therefore an exchange of energy between the mechanical system in consideration and some external agent represented by the fractionality. We shall be coming back to this point  in the explicit example we are going to analyze in Section 4. 
Constructing the functional variation of action $\delta S^{\alpha}\,=\,0\,\,,
$ we obtain after some calculus that,

\begin{equation}
{\partial L \over \partial q_i}\,-\,{d \over d{\tau}}\left({\partial L \over {\partial \dot q_i}}\right)\,-\,\left({1-\alpha\over t-\tau}\right){\partial L \over{\partial \dot q_i}}\,=\,0 ,\,\,\,\,\,i=1\cdots n \,\,.
\end{equation}

After these few words on the fractional formalism, we think that it is important to justify our choice of using the RL fractional technique instead of other very popular fractional framework: the Caputo time derivative (CPT). The latter could be another way to attack the construction of fractional DB. Nevertheless, we understood that CPT is more appropriated to applications in several engineering problems due to the fact that it has a better relation with Laplace transform.  And because the differentiation appears inside the integral.  So, it smoothes the effects of noise and numerical differentiation. Another point is that Caputo's definition is quite useful when we are treating initial problem value, but this is not our goal. For our central purpose, the RL approach is more convenient than CPT.
Having clarified our choice, we can write, from Eq. (2), that the  following integral
\begin{equation}
\label{5000}
\delta S\,=\,{1\over \Gamma(\alpha_i)}\,\delta\,\int_{\tau} \left[p \dot q\,-\,H(p_i,q_i,\tau)\right] (t-\tau)^{\alpha-1} d\tau\,=\,0
\end{equation}
so that
\begin{eqnarray}
\delta S^{\alpha}&=&{1\over \Gamma(\alpha_i)} \int_\tau [\,(\delta L)\,(t\,-\,\tau)^{\alpha-1}\,+\,L (\delta(t-\tau)^{\alpha-1})\,]\,d\tau  \nonumber \\
&=&0 \,\,,
\end{eqnarray}
where $L=p \dot q\,-\,H(p_i,q_i,\tau)$ and the rest of the calculation is as standard as the variational calculus discussed in the text books.  The modification is due to FC formalism.  However, we can readily deal with this additional factor.  Hence, after performing the variations on the Lagrangian (as in Eq. (3)) and the damping factor, and isolating the coefficients for $\delta \dot{q}$ and 
$\delta \dot{p}$ that will be equal to zero, we can write a new set of perturbed equations,
\begin{eqnarray}
\dot q_i&=&{\partial H \over{\partial \dot p_i}} \\
\dot p_i&=&-{\partial H \over{\partial \dot q_i}}\,+\,p_i\left({1-\alpha\over t-\tau}\right),
\end{eqnarray}
which can be understood as the (fractional) Hamilton-Jacobi equations when this new action functional is considered.  It is clear that when $\alpha \rightarrow 1$ our results will turn back to the usual case, as shown above. 

We showed in \cite{cresus2} that the quotient $\frac{1-\alpha}{t-\tau}\,p$ will be important in our fractional Dirac Bracket formulation.  The order $\alpha$ will be directly related to the fractional approach.
The presence of a fractional factor ${{1-\alpha}\over{t-\tau}}\,,$ is responsible for the generation of a time-dependent damping into the dynamics of the system, which is very useful to study models with smooth turbulence. Furthermore it is possible to establish a relationship between the fractional Rayleigh dissipation function and the Euler-Lagrange equation,   
\begin{equation}
\label{oito}
{\partial L \over \partial q_i}\,-\,{d \over d{\tau}}\left({\partial L \over \partial {\dot q_i}}\right)\,-\,{\partial R \over{\partial \dot q_i}}\,=\,0 ,\,\,\,\,\,i=1\cdots n ,
\end{equation}
where $R$ is the fractional Rayleigh dissipation function given by,
\begin{equation}
R\,=\,L\left({1-\alpha}\over{t-\tau}\right).
\end{equation}
Note that in Eq. (\ref{oito}), the dissipation function is part of the extended Euler-Lagrange equation.  
The origin of the third term is non-standard and is due to fractional analysis and points to a time-dependent damping factor.

\newpage
\section{Modified Variational Principle for High-Order Derivative Systems}
\subsection{Higher-Order Fractional Lagrange Equations of Motion}
Our purpose is now to extend the fractional treatment to include systems with higher derivatives present.
We shall focus on mechanical systems whose Lagrangians present second order time derivatives. The interest on acceleration-dependent Lagrangians for mechanical
systems relies on our claim that they may give us some clues on properties on field-theoretic models whose classical dynamical equations are of fourth order in
time derivative. The extension to include orders higher than two at the Lagrangian level can be readily carried out. Our fractional integral is now
\begin{equation}
\label{1}
S^{\alpha}(t)\,=\,{1\over \Gamma(\alpha_i)} \int_{\tau} L(q(\tau),\dot q(\tau),\ddot q(\tau),\tau)(t-\tau)^{\alpha-1}d\tau ,
\end{equation}
and, since we are considering all $\delta q_i$ independent, The Fractional Euler-Lagrange equations of motion can be obtained by varying the action, 
$\delta S^{\alpha}=0$.  The new fractional Euler-Lagrange equation can
be written as it follows below:
\begin{eqnarray}
\label{2}
{\partial L \over \partial q_i}&-&{d \over d{\tau}}\left({\partial L \over \partial {\dot q_i}}\right)\,+\,{d^2 \over d^2{\tau}}\left({\partial L \over \partial {\ddot q_i}}\right) \,+\,\left({1-\alpha\over t-\tau}\right){\partial L \over{\partial \dot q_i}}\,+\,\nonumber \\ 
&-&2\left({1-\alpha\over t-\tau}\right){d \over d{\tau}}{\partial L \over{\partial \ddot q_i}}\,+\,{(1-\alpha)(2-\alpha)\over (t-\tau)^2}{\partial L \over{\partial \ddot q_i}}\,=\,0 \nonumber \\
i&=&1\cdots n \nonumber\,\,. \\
\end{eqnarray}
This extended version for the fractional Euler-Lagrange equation has new time-dependent damping coefficients.
  
\subsection{Higher-Order Fractional Hamilton Equations}
We shall now consider the associated Hamiltonian formulation  for acceleration-dependent Lagrangian systems; once again, our starting point is the fractional integral (\ref{1}), for systems with n degrees of freedom.  The change of basis from $(q,\dot q, \ddot q,t)$ to the new coordinates $(q,p,\pi,t)$ is accomplished by the well-known Legendre transformation.  So, we consider a function Hamiltonian $H(q,p,\pi,t)$, and the fractional integral can be written now as 
\begin{equation}
\label{3}
S^{\alpha}(t)\,=\,{1\over \Gamma(\alpha_i)} \int_{\tau} \left[\sum_i{p_i \dot q_i}\,+\,\sum_j {\pi_j \ddot q_j}\,-\,H(p_i, \pi_i,q_i,\dot q_i,t)\right](t-\tau)^{\alpha-1}d\tau \,.
\end{equation}

Applying the variational principle yields the higher-order Hamilton equations as cast below: 
\begin{eqnarray} 
\label{hamilton eq}
\dot q_i &=& {{\partial H}\over{\partial p_i}}\,,  \\
\dot p_i &=& -{{\partial H}\over{\partial q_i}}+ \left({1-\alpha\over t-\tau}\right)p_i\,, \\ 
\ddot q_i&=& {{\partial H}\over{\partial \pi_i}}\,, \\
\dot \pi_i &=& -{{\partial H}\over{\partial \ddot q_i}}+ \left({1-\alpha\over t-\tau}\right)\pi_i \,.
\end{eqnarray}

\section{A working example: Higher-Derivative Harmonic Oscillator}
\paragraph*{} Our present task consists in analyzing a second derivative generalization of the harmonic oscillator in the version examined by Pais and Uhlenbeck \cite{pais}. We are however re-assessing it in the fractional integral framework. The classical and known Lagrangian is given by
\begin{equation}
L=\beta\overset{\cdot\cdot}{q}^{2}+\dfrac{m}{2}\overset{\cdot}{q}^{2}%
-\dfrac{m\omega^{2}}{2}q^{2}
\end{equation}
 where $m$ is the particle mass, $\omega$ is the frequency and $\beta$ is an arbitrary parameter with mass dimensional equal to (-2). In the Pais-Uhlenbeck model, $\beta$ assume the value $-mg/(2\omega^2)$, where $g$ is a small positive number ($0<g<1/4$) playing the role of a coupling constant.
\paragraph*{} The fractional Euler-Lagrange equation obtained from the Lagrangian (1) can be read below:
\begin{equation}
\label{de}
q^{\left(  4\right)  }\left(  \tau\right)  -\dfrac{2\left(  1-\alpha\right)
}{\left(  t-\tau\right)  }q^{\left(  3\right)  }\left(  \tau\right)  +\left[
\dfrac{\left(  1-\alpha\right)  \left(  2-\alpha\right)  }{\left(
t-\tau\right)  ^{2}}-\dfrac{m}{2\beta}\right]  \overset{\cdot\cdot}{q}\left(
\tau\right)  +\dfrac{m}{2\beta}\dfrac{\left(  1-\alpha\right)  }{\left(
t-\tau\right)  }\overset{\cdot}{q}\left(  \tau\right)  -\dfrac{m\omega^{2}%
}{2\beta}q\left(  \tau\right)  =0
\end{equation}
\paragraph*{} Obviously, if $\alpha=1$, we recover the known integer Pais-Uhlenbeck equation of motion. Unfortunately, we could not get an analytic solution to Eq. (2). Therefore, the remainder of our discussion shall be carried out using numerically methods with the purpose of studying its dynamical behavior and thus establish comparisons between the results of the fractional-order systems and their integer-order counterparts. Since $t$ in Eq. (2) is the time for some observer in a particular reference frame, without loss of generality, we can set $t=0$.
\paragraph*{} We go one step further appropriately choosing initial conditions, namely, 
\begin{equation}
q\left(  \tau=10^{-4}\right)  =1
\end{equation}
\begin{equation}
\overset{\cdot}{q}\left(  \tau=10^{-4}\right)  =1
\end{equation}
\begin{equation}
\overset{\cdot\cdot}{q}\left(  \tau=10^{-4}\right)  =\overset{\cdot\cdot
\cdot}{q}\left(  \tau=10^{-4}\right)  =0
\end{equation}
\paragraph*{} The results of the numerical simulations of  (\ref{de}) are plotted below, with different values of $\alpha $ and $g$. The graph in the Figure 1 compares the curves obtained with $\alpha\leq1$ and the one in Figure 2 compares the curves obtained with $\alpha\geq1$. Likewise, Figure (3) depicts the behavior of Eq. (2)  for values of $g$. In Figure (3), we set the value of $\alpha$ and plot different solutions of Eq. (2) corresponding to the different values of $g$. On the other hand, for values of $g>0.25$, the solution of Eq. (2) behaves quite differently from the previous ones. Figures (4) and (5) illustrate this fact. 
\begin{figure}[htbp]
  \centering
  \begin{minipage}[b]{0.47\textwidth}
    \includegraphics[width=\textwidth]{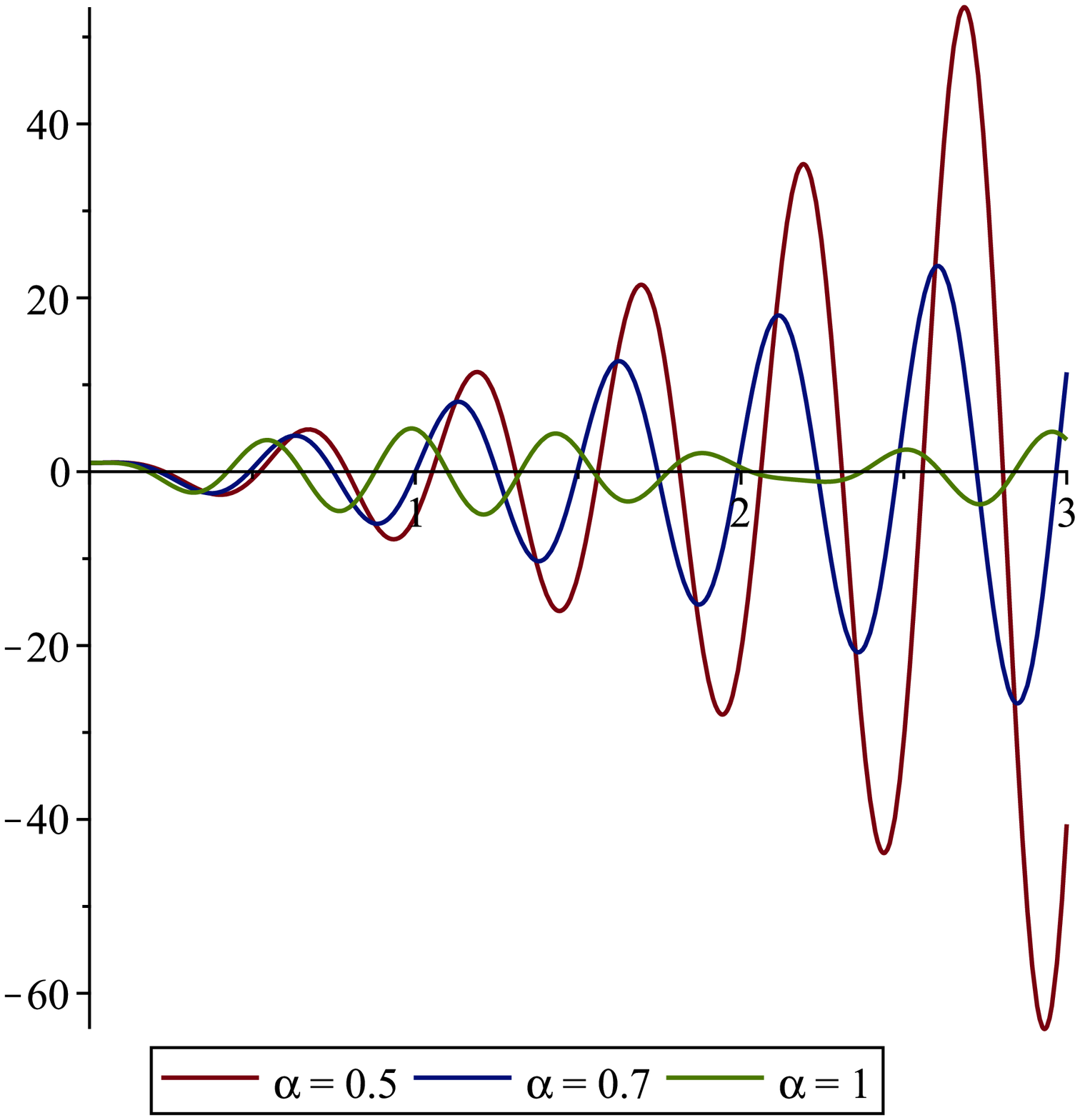}
    \caption{The running of $q(\tau)$, with $m=1, \omega=10, g=0.24$. Values of $\alpha\leq1$.}
  \end{minipage}
  \hfill
  \begin{minipage}[b]{0.47\textwidth}
    \includegraphics[width=\textwidth]{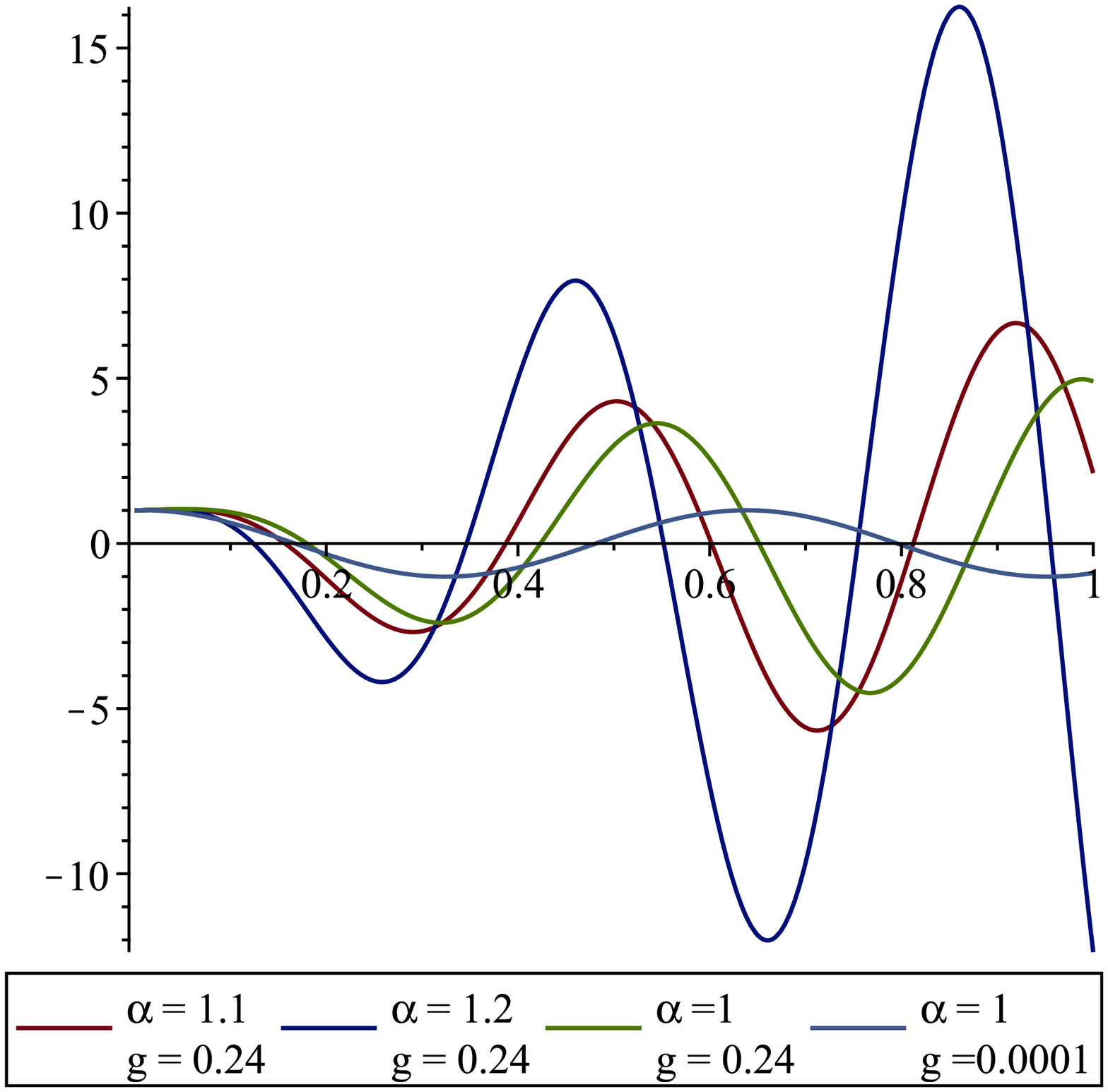}
    \caption{The running of $q(\tau)$, with $m=1, \omega=10, g=0.24$. Values of $\alpha\geq1$.}
  \end{minipage}
\end{figure}
\begin{figure}[htbp]
  \centering
  \begin{minipage}[b]{0.47\textwidth}
    \includegraphics[width=\textwidth]{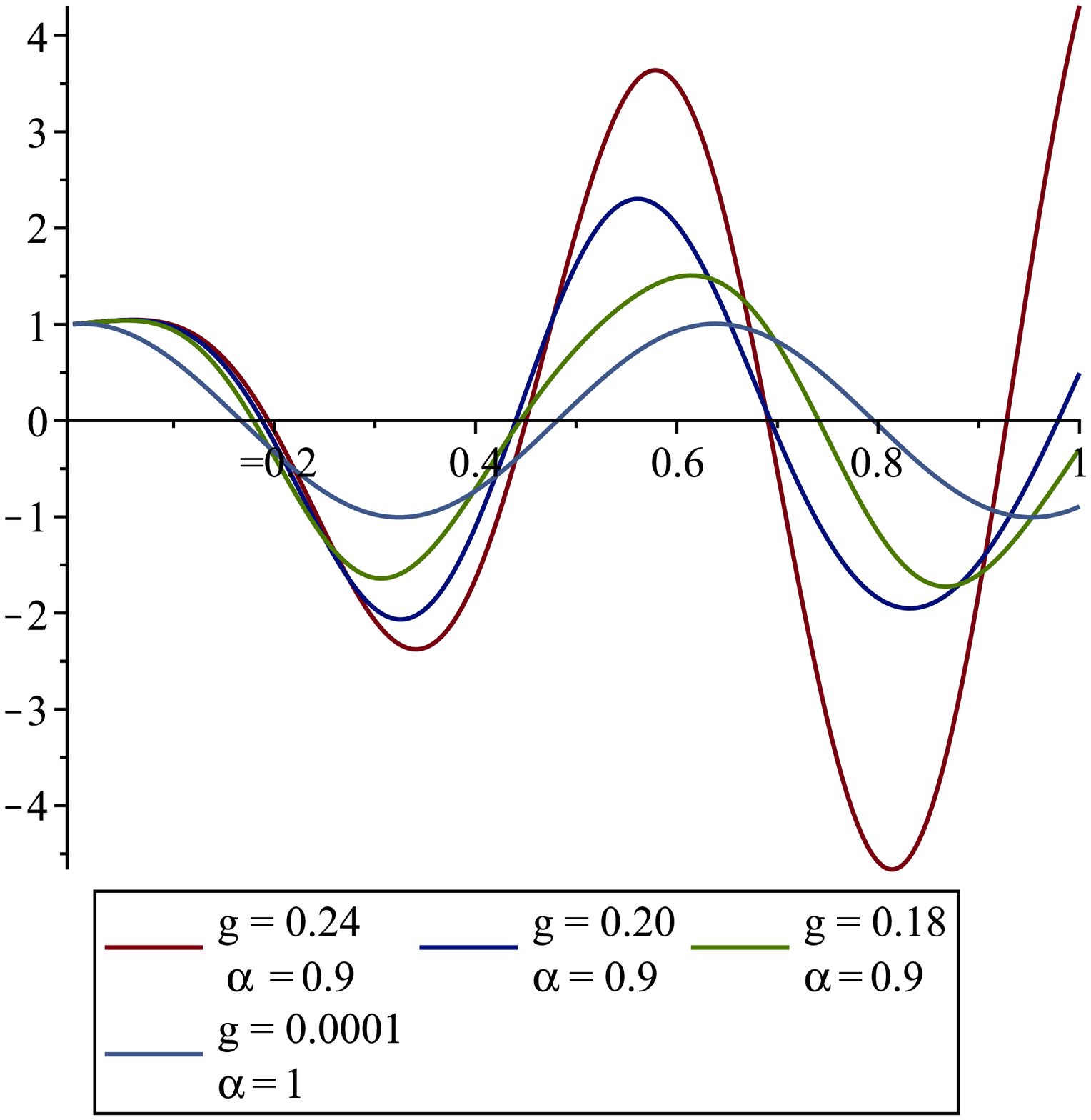}
    \caption{The running of $q(\tau)$, with $m=1, \omega=10, \alpha=0.9$. $0<g<0.25$.}
  \end{minipage}
  \hfill
  \begin{minipage}[b]{0.47\textwidth}
    \includegraphics[width=\textwidth]{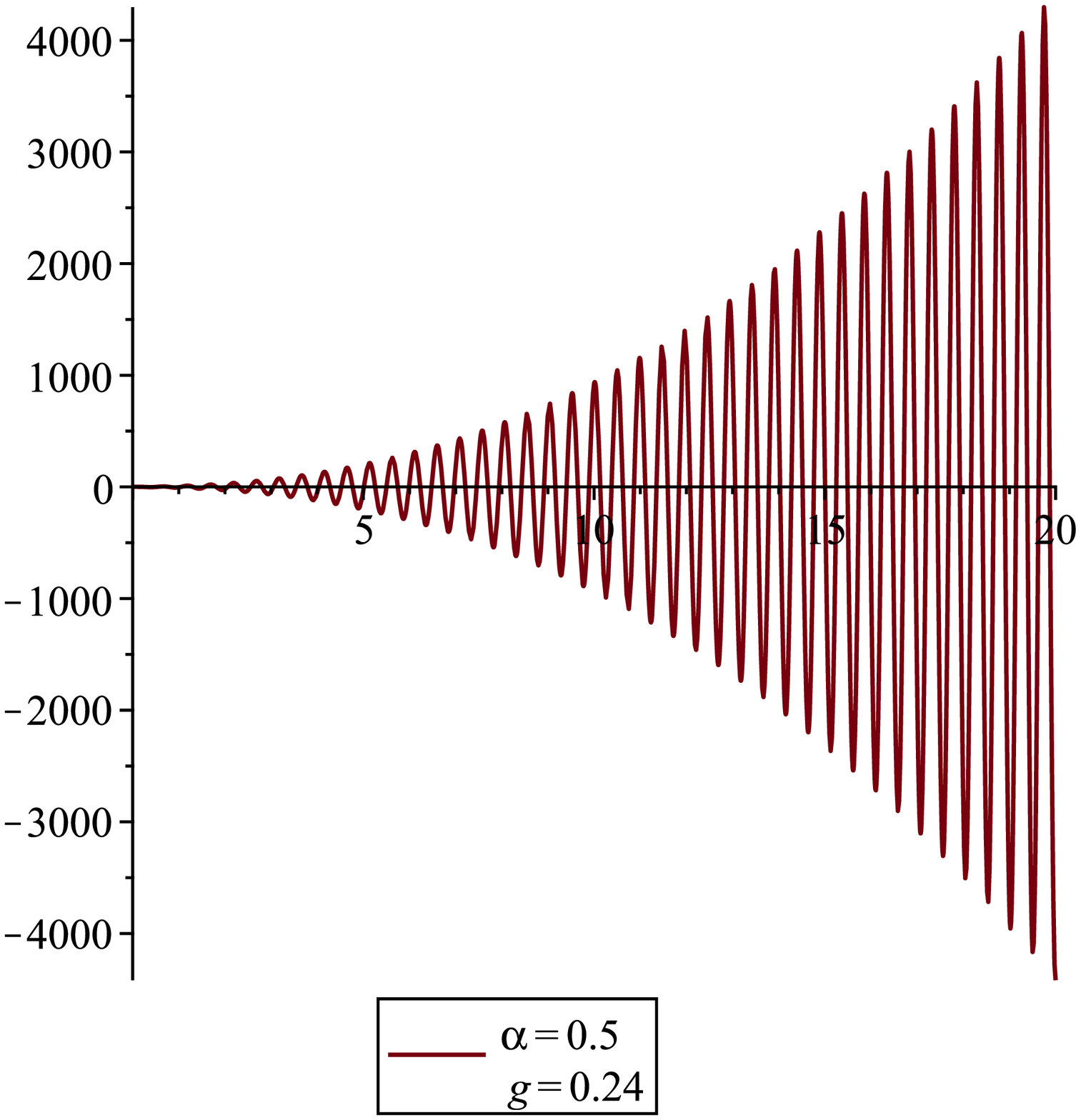}
    \caption{The running of $q(\tau)$, with $m=1, \omega=10, \alpha=0.5$. $g=0.24$.}
  \end{minipage}
\end{figure}

\begin{figure}[htpb]
\centering 
\includegraphics[width=0.47\textwidth]{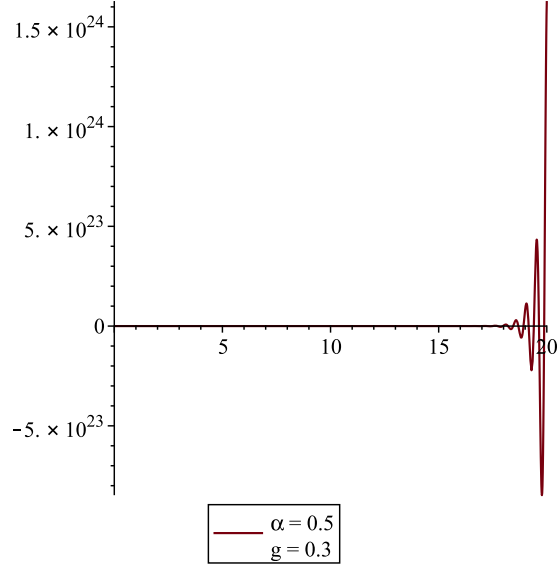}
\caption{The running of $q(\tau)$, with $m=1, \omega=10, \alpha=0.5$. $g=0.3$}
\end{figure}
\newpage
\paragraph*{} It has already been noted that the amplitude of oscillation of the system increases with time. Thus, it is also our interest to discuss this behavior by focusing on the Hamiltonian, which now should not be conserved, for the Lagrangian, at in the fractional regime, displays an explicit time dependence thorough the time-dependent weight factor. By writing the fractional action integral (\ref{1}) for Lagrangian (1), we obtain

\begin{equation}
S^{\alpha}\left[  q\right]  \left(  t\right)  =\dfrac{1}{\Gamma\left(
\alpha\right)  }%
{\displaystyle\int}
\left(  \beta\overset{\cdot\cdot}{q}^{2}+\dfrac{m}{2}\overset{\cdot}{q}%
^{2}-\dfrac{m\omega^{2}}{2}q^{2}\right)  \left(  t-\tau\right)  ^{\alpha
-1}d\tau
\end{equation}
\paragraph*{} Let us write down the following action
\begin{equation}
S^{^{\prime}\alpha}\left[  q\right]  \left(  t\right)  =\dfrac{1}%
{\Gamma\left(  \alpha\right)  }%
{\displaystyle\int}
\left(  \gamma C^{2}+\overset{\cdot\cdot}{q}C+\dfrac{m}{2}\overset{\cdot}%
{q}^{2}-\dfrac{m\omega^{2}}{2}q^{2}\right)  \left(  t-\tau\right)  ^{\alpha
-1}d\tau
\end{equation} 
where $C$ is a new coordinate which depends on the variable $\tau$; $\gamma$ is a constant.
\paragraph*{} The important point to remember here is that two different Lagrangians are classically equivalent once they can transformed one another by local (in the time-like sense) coordinate redefinitions. By taking the action (7), we can readily obtain the equation of motion for the $C$-coordinate, which reads
\begin{equation}
C=-\dfrac{1}{2\gamma}\overset{\cdot\cdot}{q}
\end{equation} 
\paragraph*{} Here, it is worthwhile to recall that, since $C$ behaves as an auxiliary coordinate (its dependence is purely algebraic), it is a perfectly legitimate procedure to replace it in the action (6) by its equation of motion (8). After this substitution, by comparing the actions (6) and (7), one gets the action given below:
\begin{equation}
S^{^{\prime}\alpha}\left[  q\right]  \left(  t\right)  =\dfrac{1}%
{\Gamma\left(  \alpha\right)  }%
{\displaystyle\int}
\left(  -\dfrac{1}{4\beta}C^{2}+\overset{\cdot\cdot}{q}C+\dfrac{m}{2}%
\overset{\cdot}{q}^{2}-\dfrac{m\omega^{2}}{2}q^{2}\right)  \left(
t-\tau\right)  ^{\alpha-1}d\tau
\end{equation}
By integrating the second term by parts yields
\begin{equation}
S^{^{\prime}\alpha}\left[  q\right]  \left(  t\right)  =\dfrac{1}%
{\Gamma\left(  \alpha\right)  }%
{\displaystyle\int}
\left[  -\dfrac{1}{4\beta}C^{2}-\overset{\cdot}{q}\overset{\cdot}{C}-C\left(
\dfrac{1-\alpha}{t-\tau}\right)  \overset{\cdot}{q}+\dfrac{m}{2}\overset{.}%
{q}^{2}-\dfrac{m\omega}{2}q^{2}\right]  \left(  t-\tau\right)  ^{\alpha
-1}d\tau
\end{equation}
\paragraph*{} We now consider the transformation
\begin{equation}
B=\sqrt{m}q-\dfrac{C}{\sqrt{m}}
\end{equation}
where $B$ is a new coordinate. Replacing eq. (11) in the action (10), the Lagrangian can be explicitly expressed only in terms of the new basis of coordinates, $B$ and $C$, namely
\begin{equation}
L^{^{\prime}}=\dfrac{1}{2}\overset{\cdot}{B}^{2}-\dfrac{1}{2m}\overset{\cdot
}{C}^{2}-\dfrac{1}{2}\omega^{2}B^{2}-\dfrac{1}{4}\left(  \dfrac{1}{\beta
}+\dfrac{2\omega^{2}}{m}\right)  C^{2}-\dfrac{\omega^{2}}{\sqrt{m}}BC-\left(
\dfrac{1-\alpha}{t-\tau}\right)  \dfrac{1}{\sqrt{m}}\overset{\cdot}{B}%
C-\dfrac{1}{m}\left(  \dfrac{1-\alpha}{t-\tau}\right)  C\overset{\cdot}{C}
\end{equation}
which shows it is manifestly time-dependent. The steps above have been included here to explicitly show why the Hamiltonian is not conserved. This happens due to the explicit time dependence of the Lagrangian rewritten after the coordinate redefinitions.

\section{Discussion and Final Considerations}
\paragraph*{} In this paper, we have applied the Riemann-Liouville approach and the fractional Euler-Lagrange equations to obtain the fractional non-linear dynamics equation involving Lagrangians with derivatives of higher order. In particular, we have contemplated the example of the Pais-Uhlenbeck harmonic oscillator. The results presented in Section 4 points to interesting aspects of the interplay between fractionality and higher derivatives.
\paragraph*{} At first, let us consider the case where $\alpha<1$ and $g$ are fixed. Surprisingly, Figure (1) shows that the system maintains its oscillatory characteristics  and its amplitude seen to grow indefinitely, in contrast to the Pais-Uhlenbeck solution, which has a variable amplitude, but with fixed maximum value. For example, the smaller the value of $\alpha$, the bigger the amplitude. For $\alpha =1$ and in the limit with $g\rightarrow0$, the system displays the expected behavior for a simple harmonic oscillator. On the other hand, for values of $\alpha>1$, Figure (2) shows that the system oscillates with an amplitude that grows with fractionality. 
\paragraph*{} We now switch to the case with fixed $\alpha$ and $g$ in the interval $\left(0,0.25\right)$. Analysing Figure (3), it is possible to notice that the amplitudes of oscillations grow whenever $g$ increases. Finally, for values of $g>0.25$, not only the amplitudes increase drastically, but the system looses its oscillatory behavior, as shown in the Figure (5). At this point, we wish to stress the difference in the behavior of the system for values of $g$ greater to $0.25$if compared with values smaller than $0.25$. Figures (4) and (5) illustrate this fact. 
\paragraph*{} The main statement of this paper is that the system governed by Eq.(20) exhibits an oscillatory motion with increasing amplitudes, although this equation presents a new time-dependent damping coefficient.
\paragraph*{} To conclude, we re-inforce that the motivation for our work relies on the relationship between the physics of mechanical models as a one-dimensional reduction of corresponding field-theoretic models, so that properties of the former may help in understanding some features of the latter. And, in considering the fractional dynamics of higher-derivative mechanical models, we are opening up a path to discuss the fractionality regime of higher-derivative field theories. We are presently working on the field-theory versin and we shall soon report on that elsewhere.

\end{document}